\documentclass[12pt,preprint]{aastex}
\shorttitle{Substructure in clusters of galaxies}
\shortauthors{Natarajan \& Springel}

\newcommand{\msun}{{\rm M}_\odot}

\begin{document}

\title{Abundance of substructure in clusters of galaxies}

\author{Priyamvada Natarajan\altaffilmark{1} and  Volker Springel\altaffilmark{2}}

\altaffiltext{1}{Department of Astronomy, Yale University, P.O. Box 208101, 
       New Haven, CT 06520-8101, USA, priya@astro.yale.edu}
\altaffiltext{2}{Max-Planck Institut f\"ur Astrophysik, Garching, Germany, 
            volker@mpa-garching.mpg.de}

\begin{abstract}
It is widely recognized that cold dark matter models predict abundant
dark matter substructure in halos of all sizes. Galaxy-galaxy lensing
provides a unique opportunity to directly measure the presence and the
mass of such substructures in clusters of galaxies. Here we present
the mass function of substructures obtained from lensing in massive
HST cluster lenses, and compare it with that obtained from
high-resolution cosmological N-body simulations. We find excellent
agreement in the slope and amplitude in the mass range
$10^{11}\,-10^{12.5}\,{\rm M}_{\odot}$ probed by the observations,
highlighting a significant success of the CDM theory. At lower mass
scales, the simulations predict a large abundance of small
substructures below our detection threshold with galaxy-galaxy
lensing, with many of them being plausibly associated with faint
cluster galaxies. Our results suggest that the CDM substructure
abundance on the scale of clusters is in good agreement with present
observational data.
\end{abstract}

\keywords{cosmology: dark matter --- gravitation: gravitational lensing}

\section{Introduction}

Arguably the strongest challenges presently faced by the otherwise
highly successful cold dark matter (CDM) paradigm for structure
formation in the Universe occur on `small scales', i.e.~within
individual dark matter halos. One of the most prominent of these
problems is the long standing and controversially debated question
whether rotation curves of low surface brightness galaxies can be
accommodated in the CDM theory
\citep[e.g.][]{McGaugh1998,Hayashi2003,Navarro2004}. Another important
issue is concerned with the amount of dark matter substructure
theoretically predicted by CDM, and whether this is in conflict with
the relative paucity of satellites observed in the Milky Way and the
Local Group \citep[e.g.][]{Moore1999,Klypin1999,Stoehr2002}.

Interestingly, both of these problems have only been recognized once
N-body simulations reached high enough resolution for detailed studies
of the phase-space structure of dark matter halos, allowing them to
probe the inner dark matter cusp to sub-kpc scales, and to establish
that dark halos are filled with a host of self-bound dark matter
satellites, instead of being the smooth halos envisioned based on
earlier low-resolution work. It has now become clear that the
existence of substructure is a generic prediction of hierarchical
structure formation in CDM models, where the assembly of collapsed
mass proceeds via a merger hierarchy that allows some of the infalling
dense lumps to survive as dynamically distinct substructure inside
virialized halos until late times.

On the scales of ordinary galaxies like the Milky Way
($\sim\,10^{12}\,{\rm M}_{\odot}$), the number of bound satellites
predicted by the simulations is much larger than the dozen or so
satellites actually detected around our Galaxy.  Comparing the observed
velocity distribution function of these satellites with the predicted
one \citep{Moore1999,Klypin1999}, this has been interpreted as
evidence for a serious surfeit of dark matter satellites, and even
considered to mark a ``crisis'' of the $\Lambda$CDM model.  However,
as \citet{Stoehr2002} have shown, there is significant uncertainty in
the conversion of stellar line-of-sight velocity dispersion into peak
circular velocities of dark matter substructures. Using high-resolution
simulations of a Milky Way sized halo, \citeauthor{Stoehr2002} in fact
were able to show that all the known satellites of the Milky Way can
be comfortably hosted by the most massive dark matter substructures
expected for its halo.  In this picture, there is however still a sea
of small mass dark matter sub-halos which are largely devoid of stars.
The leading hypothesis to explain this situation is to allude to
baryonic processes of galaxy formation, inhibiting star formation
preferentially in low mass halos. Proposed processes for such feedback
include a photo-ionizing UV background or the expulsion of gas in
shallow potential wells by supernova explosions
\citep[e.g.][]{Bullock2000,Benson2002,Kravtsov2004}.

If vast numbers of {\em dark} sub-halos exist, lensing might be the
best way to detect them \citep[e.g.][]{Trentham2001}. This is not only
true for the scales of galaxies, but particularly for rich clusters of
galaxies, where strong and weak lensing effects can be combined in a
powerful way to construct detailed mass models for clusters. This
allows in principle a direct mapping of the sub-halo mass function,
thereby providing an important test of the CDM paradigm. Upon
comparison with the observed galaxy luminosity function of clusters,
it can also give valuable insights into the galaxy formation process.

\begin{table*}
\begin{center}
\begin{tabular}{lcccccccc}
\hline\hline\noalign{\smallskip}
${\rm Cluster}$&{z} & ${\sigma_{0\ast}}$&${r_{t^\ast}}$&${M_{\rm ap}/L_v}$&$\rm {M^\ast}$&
$\sigma_{\rm clus}$ & ${\rho_{\rm clus}(r = 0)}$\\
& & (km\,s$^{-1}$) & (kpc) & 
(M$_\odot$/L$_\odot$) & (10$^{11}$M$_\odot$) & (km\,s$^{-1}$) &
(10$^6$ $\msun$ kpc$^{-3}$)\\
\noalign{\smallskip}
\hline
\noalign{\smallskip}
{A\,2218} & ${0.17}$ & ${180\pm10}$ & ${40\pm12}$ &
${5.8\pm1.5}$ & $\sim\,14 $ & ${1070\pm70}$  &  {3.95}\\

{A\,2390} & ${0.23}$ & ${200\pm15}$ & ${18\pm5}$ &
${4.2\pm1.3}$ & $\sim\,6.4 $  &${1100\pm80}$& {16.95}\\

{AC\,114}  & ${0.31}$ & ${192\pm35}$ & ${17\pm5}$ &
${6.2\pm1.4}$ & $\sim\,4.9 $ &${950\pm50}$& {9.12}\\ 

{Cl\,2244$-$02} & ${0.33}$ & ${110\pm7}$ & ${55\pm12}$ &
${3.2\pm1.2}$ & $\sim\,6.8 $ &${600\pm80}$  & {3.52}\\

{Cl\,0024+16} & ${0.39}$ & ${125\pm7}$ & ${45\pm5}$ &
${2.5\pm1.2}$ & $\sim\,6.3 $ &${1000\pm70}$ & {3.63}\\

{Cl\,0054$-$27} & ${0.58}$& ${230\pm18}$ & ${20\pm7}$ &
${5.2\pm1.4}$ & $\sim\,9.4 $ &${1100\pm100}$ & {15.84}\\
\noalign{\smallskip}
\hline
\end{tabular}
\end{center}
\end{table*}

 In this letter, we present results of the first comparison between the
lensing determined substructure mass function in clusters on mass
scales ranging from about $10^{11}\,-10^{12.5}\,{\rm M}_{\odot}$ with
that inferred from high resolution cosmological N-body simulations.
To this end, we construct high resolution mass maps of galaxy clusters
by applying galaxy-galaxy lensing techniques where substructure inside
clusters is detected and mapped using the anisotropies that they
produce in the observed shear field.  We compare our lensing
measurements directly with results from high-resolution N-body
simulations, allowing us to test the robustness of the CDM model and
the associated hierarchical galaxy formation paradigm. Note that
compared to the scales of galaxies, we expect many more dark matter
structures to be visible optically in clusters, making the comparison
of sub-halo mass functions less affected by uncertainties in the
galaxy formation physics.  Therefore, full consistency between the
abundance of optically detected galaxies, substructure in CDM N-body
models, and sub-halos detected by lensing can be asked for;
establishing such a consistency can be viewed as a strong test of the
theoretical paradigm.  Below, we outline our methodology for mapping
substructure in clusters. We then present the mass function derived
from the maximum-likelihood analysis. This is followed by a comparison
with the substructure detected in simulated clusters, and a discussion
of our results. We adopt $h = 0.7$, $\Omega_0 = 0.3$ and
$\Omega_{\Lambda} = 0.7$.

\section{Quantifying substructure in clusters with gravitational lensing}

We obtain the mass spectrum of clumps in a cluster by combining
constraints from strong lensing observations (highly magnified
multiply imaged systems) and weak lensing (the radially averaged
tangential shear profile). To this end, we use a self-similar,
parametric mass model to describe the cluster as a composite of a
large-scale smooth mass distribution and several sub-clumps
\citep{Natarajan1997,Natarajan1998,Natarajan2002,Natarajan2004}.
These sub-halos are associated with bright, early-type galaxies in the
cluster under the assumption that mass traces light. The local
anisotropies in the shear field induced by the sub-halos in their
vicinity is then used statistically to quantify the mass of the
sub-clumps and their spatial extents. A likelihood method is used to
retrieve characteristic halo parameters
\citep[e.g.][]{Natarajan1997,Natarajan1998,Geiger1998}. On
applying these techniques to an ensemble of HST cluster lenses
(results are presented in Table~1) we find
that the spatial extents inferred are consistent with tidal stripping;
early-type galaxies do possess dark halos that extend well beyond the
light but these halos are more compact than those around field
galaxies of equivalent luminosity.

In performing the likelihood analysis to obtain characteristic
parameters for the sub-clumps in the cluster we assume that light
traces mass. This is an assumption that is well supported by
galaxy-galaxy lensing studies in the field \citep{Wilson2001} as well
as in clusters \citep{Clowe2002,Hoekstra2003}. The individual galaxies
and the smooth cluster component are modeled self-similarly with
truncated pseudo-isothermal mass distributions (PIEMD). The parameters
that characterize a truncated PIEMD are: a truncation radius $r_t$
identified with the tidal radius, a core radius $r_0$ and a central
velocity dispersion $\sigma_0$. For the smooth component the values of
these parameters are set by the observed strong lensing features, and
for the galaxies, combined constraints from the strong lensing and the
weak shear field determine the best-fit parameters for a fiducial
galaxy halo. These values recovered from the maximum likelihood
analysis are shown in Table~1.  In order to relate the observed light
distribution in the early-type cluster galaxies to their masses a set
of physically motivated scaling laws are assumed:
\begin{eqnarray}  
\sigma_0 = \sigma_{0*}\,\left(\frac{L}{L*}\right)^{1/4};\;\;r_0 =
r_{0*}\,\left(\frac{L}{L*}\right)^{\alpha};\;\; r_t = r_{t*}\,\left(\frac{L}{L*}\right)^{\alpha}.
\end{eqnarray}
The total mass of the sub-halo associated with a galaxy of luminosity 
$L$ is:
\begin{eqnarray}  
M \propto \sigma_{0*}^2 r_{t*} \left(\frac{L}{L*}\right)^{1/2+\alpha}; \;\;\;
\frac{M}{L} \propto \sigma_{0*}^2 r_{t*} \left(\frac{L}{L*}\right)^{1/2-\alpha}.
\end{eqnarray}
Note however that for our choice of mass model and value adopted for
the exponent $\alpha$, the mass to light ratio is not constant with 
radius within an individual galaxy halo. The derived mass spectrum of
sub-halos is not a strong function of $\alpha$, as discussed in 
\citep{Natarajan2004}. 

Dark halos are associated with the locations of bright, early-type
cluster galaxies and the fiducial parameters for a typical halo are
then extracted from the likelihood analysis. A high resolution mass
model for the entire cluster is built using the strong lensing regime
to constrain the inner region ($r\,\sim\,r_{\rm Einstein}$), and the
local anisotropies in the shear field are used to obtain properties of
the galaxy halos around early-type cluster members.  Since the
procedure involves a scaled, self-similar mass model that is
parametric, we obtain a mass estimate for the dark halos (sub-clumps)
of the cluster galaxies as a function of their luminosity. This
provides us with a clump mass spectrum. Note that tidal truncation by
the cluster causes these halo masses to be lower than that of
equivalent field galaxies at comparable redshifts obtained from
galaxy-galaxy lensing. The fraction of mass in the clumps is only
10-20\% of the total mass of the cluster within the inner
$500\,h^{-1}\,{\rm kpc}$ of these high central density clusters.  We
are limited to this spatial scale as the lensing analysis was
performed on HST-WFPC2 data with pointings at cluster centers.  The
remaining 80-90\% of the cluster mass is consistent with being
smoothly distributed (in lumps with mass $M\,<\,10^{10}\,{\rm
M}_{\odot}$).

\begin{figure*}
\begin{center}
\resizebox{13cm}{7.5cm}{\includegraphics{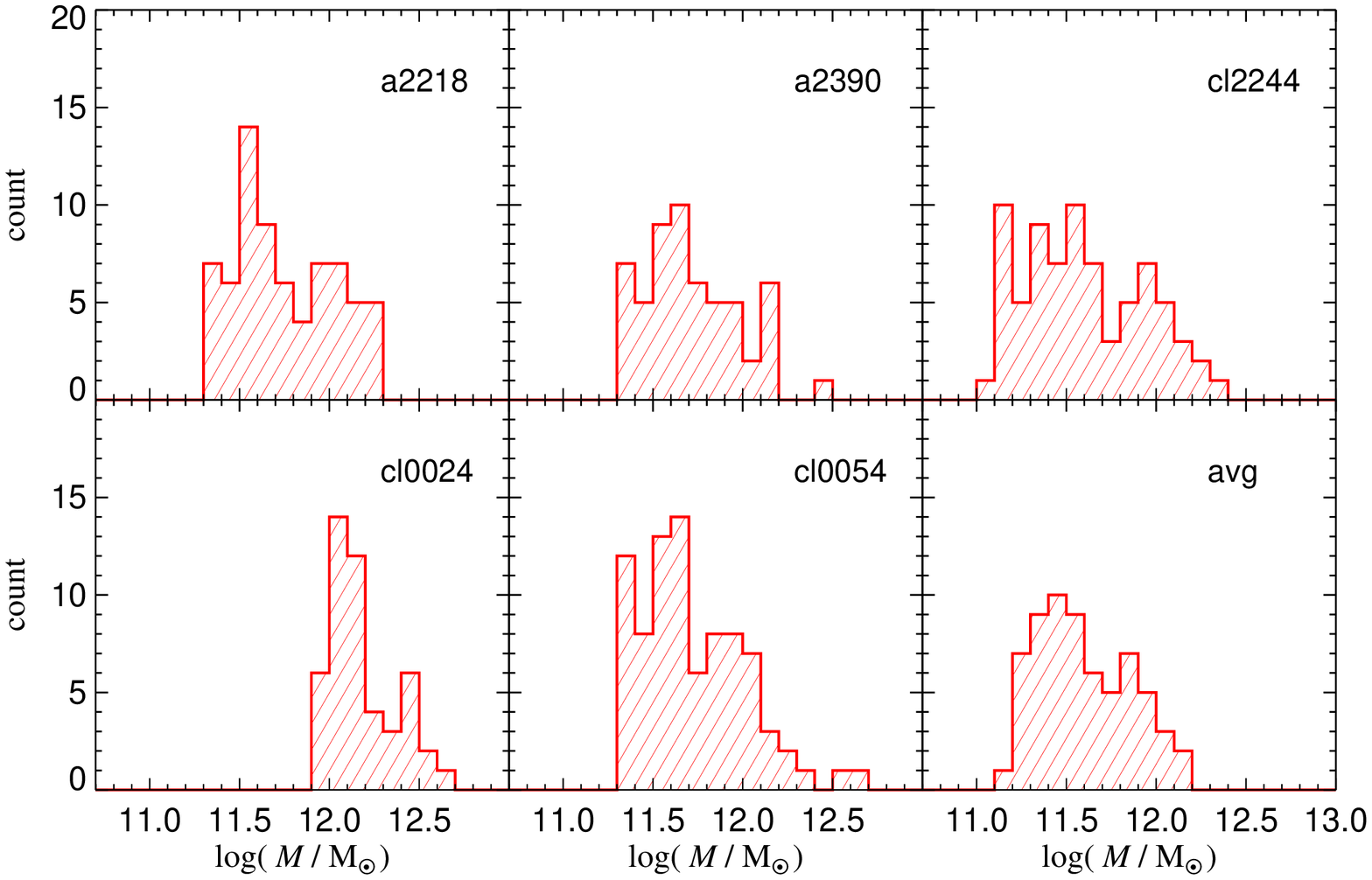}}%
\end{center}
\caption{The substructure mass function retrieved from galaxy-galaxy
lensing for 5 massive HST cluster-lenses. This mass spectrum is
constituted by the dark halos associated with the 40 brightest
galaxies in these clusters. The panel on the bottom right shows the
coadded and averaged clump spectrum of clusters A2218, A2390, and Cl0054.
\label{fig1}}
\end{figure*}

In Fig.~\ref{fig1}, we show the mass function retrieved from
galaxy-galaxy lensing for each of the five HST clusters. There is a low-mass
cut-off in the observed clump spectrum, at around $10^{11}{\rm
M}_\odot$, which is due to observational limitations. The mass
resolution of this technique is limited by the depth and field of view
of the Wide Field Planetary Camera (WFPC2) aboard the Hubble Space
Telescope, by the number of background galaxies per foreground lens,
and the reliability with which shapes can be measured for the faintest
background galaxies in the HST image of the central region.
Unfortunately, this limits the number of reliably determined lumps per
rich cluster to about 40, implying that only the massive end of the
clump spectrum corresponding to the brightest cluster galaxies can be
probed. However, despite the low number statistics, the individual
cluster measurements show a marked rise in substructure abundance
towards lower mass scales. This becomes particularly apparent once the
clusters are stacked, as we have done in the lower right panel of
Fig.~\ref{fig1}. 

\section{Cluster substructure in high resolution $\Lambda$CDM simulations}

We analyze substructure in high-resolution dark matter simulations of
clusters formed in the $\Lambda$CDM model. Our set of simulations is
taken from the study carried out by \citet{Springel2001} of
a single rich cluster of mass $8.4\times 10^{14}\,h^{-1}{\rm
M}_\odot$, simulated in 4 steps of ever increasing resolution using
the parallel tree-code {\small GADGET} \citep{gadget2001}. In these
simulations (referred to as `S1' to `S4'), the particle mass
resolution increases from $6.87\times 10^9$ to $4.68\times
10^7\,h^{-1}{\rm M}_\odot$, corresponding to $1.3\times 10^5$
particles up to about 20 million within the virial radius, making the
highest resolution simulation in this series one of the best resolved
simulations of a single rich cluster carried out to date. This
simulated cluster has a comparable central density and mass within the
inner $500\,h^{-1}\,{\rm kpc}$ as the massive HST cluster-lenses
studied here.

Self-bound gravitational substructure is found with the algorithm
{\small SUBFIND} \citep{Springel2001}. It starts by determining
locally over-dense dark matter substructure candidates in a fully
adaptive fashion, and then subjects each of them to a gravitational
unbinding procedure, such that a catalog of self-bound dark matter
substructures results. In the lowest resolution simulation S1 of the
series, 118 substructures can be detected, a number that increases to
more than 4600 in the high-resolution simulation S4. Note however that
these additionally resolved small substructures are all of ever lower
mass; already the low resolution simulation captures the correct
number of massive satellites, with (at least on average) the correct
mass.

The measured differential mass function of substructures appears to be
a power law, ${\rm d}N/{\rm d}m \propto m^{-\alpha}$, with slope close
to $\alpha\simeq -1.8$ \citep{Springel2001,Helmi2002,DeLucia2004}.
Interestingly, this is very close to the result by \citet{Lee2004} who
has attempted an analytic calculation of the sub-halo mass function
based on modeling the complex dynamical history of galaxies in the
cluster with a parameterized model to account for the effects of
global tidal truncation \citep[see also][]{Taylor2001}. Given that we
have also demonstrated that the spatial extents of substructures
inferred from galaxy-galaxy lensing are consistent with the tidal
stripping hypothesis, this lends support to the validity of a simple
tidal-limit approximation.  Similar to the lensing result, both the
analytic model of \citet{Lee2004} and the numerical simulations find
that only about 10\% of the sub-halo mass is bound in substructures,
most of it in a handful of most massive sub-halos which dominate the
cumulative mass in the substructures.

We note that numerical studies by \citet{DeLucia2004} find that the
substructure mass function depends only weakly on the properties of
the parent halo mass. Also, the mass fraction in substructure is
relatively insensitive to the tilt and overall normalization of the
primordial power spectrum \citep{Zentner2003}. Only for radically
altered CDM models, for example by truncating small-scale power,
\citeauthor{Zentner2003} find that their models yield projected
substructure mass fractions that are lower than the estimates from
strong lensing.

\section{Comparison}

\begin{figure*}[!htp]
\begin{center}
\resizebox{13cm}{7.5cm}{\includegraphics{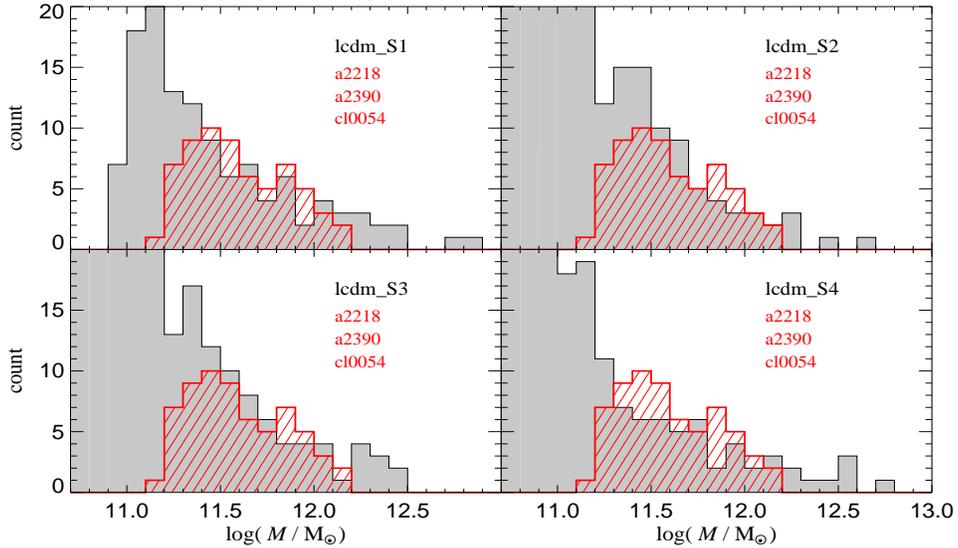}}%
\end{center}
\caption{Comparison of the substructure mass function of the stacked
  clusters A2210, A2390, and Cl0054 (shaded histogram) with N-body
  simulations of a rich cluster of galaxies (filled histogram),
  carried out at four different numerical resolutions.
\label{fig3}}
\end{figure*}

We now compare the galaxy-galaxy lensing results with the
substructure mass function obtained from the N-body simulations. In
Fig.~\ref{fig3}, we show histograms for the distribution of
substructure masses in the four simulations of \citet{Springel2001}
and contrast them individually with the stacked result for the
clusters A2218, A2390, and Cl0054. Since the number of substructures
in the observable mass range is quite small, we expect large
system-to-system variations between different clusters. The stacking
of clusters of similar mass that we applied here reduces the
associated scatter somewhat.

Comparing the lensing result with the four simulated clusters (which
also show some numerical scatter amongst each other), we find broad
general agreement in the mass range $10^{11}\,-10^{12.5}\,{\rm
M}_{\odot}$, both in the amplitude and the slope of the substructure
mass function. Given that no free parameter or scaling has been
applied to obtain this match, the agreement is in fact remarkable.

However unlike the observations, the simulations show no low-mass
cut-off. If this cut-off is entirely due to resolution effects, which
appears plausible, the interpretation is that the observations only
see the `tip of the iceberg' of the substructure distribution as far
as their number is concerned, even though they detect most of the mass
in substructures. Fig.~\ref{fig3} gives also a hint that the
simulations may systematically over-predict the mass of the most
massive satellites, but we caution that such an effect could also be
caused a systematic effect in the mass estimate based on the lensing
technique, for example if the massive substructures preferentially
correspond to halos that have fallen in most recently.

\section{Discussion}

On comparing the clump mass spectrum obtained for galaxy clusters from
high-resolution N-body simulations of $\Lambda$CDM models to those
obtained from galaxy-galaxy lensing in HST cluster lenses, we find
excellent agreement.  Despite the fact that the lensing analysis
assumes that mass traces light and is only sensitive to a restricted
mass range, it is clear that there is no substructure problem in CDM
on mass scales spanning $10^{11}\,-10^{12.5}\,{\rm M}_{\odot}$.  This
is in sharp contrast to the situation on galactic scales, where the paucity
of observed satellites when compared with the abundant dark matter
substructure predicted by simulations has been characterized as a
crises for CDM. While the severity of this problem has probably been
overstated initially -- in fact it may have partially gone away by now
\citep{Stoehr2002} -- it is clear that CDM predicts a rich spectrum of
dark matter substructure extending to very small masses, both for
clusters and galactic halos. In clusters, up to several hundred of the
most massive substructures can be directly identified with the
luminous cluster galaxies, and semi-analytic models of galaxy
formation show that this association leads to highly successful models
for the population of cluster galaxies \citep{Springel2001}. On the
other hand, on galactic scales, a much larger number of truly dark
satellites must be present according to the N-body
models. Gravitational lensing is probably our best bet to detect such
structures of small mass, and in fact, on these small scales, the
observed flux anomalies in multiply-imaged quasar systems
\citep{Mao1998,Chiba2002,Dalal2002,Metcalf2001,Metcalf2002,Mao2004} have been
interpreted as evidence for significant substructure in the mass range
$10^4\,<\,M\,<\,10^8\,{\rm M}_{\odot}$.

In this letter, we have shown that the clump mass function
independently determined from lensing (a technique that is unaffected
by the dynamical state of the cluster) is in excellent agreement with
that obtained in high resolution cosmological N-body simulations of
clusters of galaxies in the $\Lambda$CDM model.

\acknowledgements

PN thanks her collaborators on the HST cluster-lenses project: 
Jean-Paul Kneib, Ian Smail and Richard Ellis.

\bibliographystyle{apj}
\bibliography{ms}

\newpage

\newpage 

\end{document}